\documentclass[reprint,aps,prd,amsmath,showpacs,aps]{revtex4-1}
\usepackage{hyperref}
\usepackage{bm}
\usepackage{amssymb}
\usepackage{graphicx}
\begin{document}
\title{Can Galileons solve the muon problem?}

\author{Henry Lamm}
\email{hlammiv@asu.edu}

\affiliation{Physics Department, Arizona State University, Tempe, Arizona 
85287, USA}
\date{\today}

\begin{abstract}
The leptonic bound states positronium and muonium are used to constrain Galileon contributions to the Lamb shift of muonic hydrogen.  Through the application of a variety of bounds on lepton compositeness, it is shown that either the assumption of equating the charge radius of a particle with its Galileon scale radius is incompatible with experiments or the scale of Galileons must be $M>1.33$ GeV, too large to solve the muon problem.  The possibility of stronger constraints in the future from true muonium is discussed.
\end{abstract}

\maketitle
\section{Introduction}
\label{sec:1}
Several measurements in muon physics\cite{PhysRevD.73.072003,Antognini:1900ns,Pohl:2010zza,Aaij:2014ora,CMS:2014hha} have varying levels of disagreement with 
theoretical calculations.  This ``muon problem'' could be a sign of the violation of \textit{lepton universality} from beyond 
standard model (BSM) physics.  That the muon should be more susceptible to new physics is intuitively reasonable from an effective field theory point of view, given its dramatically larger mass compared to the electron, which can lead to observables proportional to powers of $m_l/\Lambda$, where $\Lambda$ is the cutoff scale.  This is similar to the enhancement of weak interactions in muonic systems discussed in Ref.~\cite{PhysRevD.91.073008}.  While larger effects with the $\tau$ or quarks are not impossible, the relative simplicity of muonic physics (little QCD contamination, longer lifetime) makes signals easier to detect.  In this paper we will consider two facets of the muon problem: the anomalous magnetic moment of the muon $a_\mu$, and the proton charge radius from muonic hydrogen $(r_P)_\mu$.

 Over the years, a number of ideas incorporating BSM physics have been proposed to solve the muon problem \cite{Heeck:2010pg,Jaeckel:2010xx,Batell:2011qq,Batell:2009jf,TuckerSmith:2010ra,Barger:2010aj,Barger:2011mt,Carlson:2012pc,Onofrio:2013fea,Wang:2013fma,Kopp:2014tsa,Gomes:2014kaa,Karshenboim:2010cm,Karshenboim:2010cg,Karshenboim:2010cj,Karshenboim:2010ck,Karshenboim:2011dx,Karshenboim:2014tka}.  A recent proposal that disformal scalars \cite{Brax:2014vva} arising from Galileon theories \cite{Nicolis:2008in} could resolve the $r_P$ discrepancy\cite{Brax:2014zba} offers the tantalizing possibility that gravitational effects could be at play in atomic systems.   We devote the work of this paper to investigating the viability of this model in the leptonic bound states positronium $(e^+e^-)$ and muonium, $(\mu^+e^-)$.  Additionally, we will make speculations about the possible limits that a measurement of true muonium $(\mu^+\mu^-)$ might set.  While this system is yet to be detected, near-term experiments \cite{Celentano:2014wya,Benelli:2012bw} are planned to 
possibly detect and characterize it.  Due to its larger reduced mass, true muonium (TM) has novel features compared to positronium (Ps) and muonium (Mu) \cite{Brodsky:2009gx,PhysRevD.91.073008}, including the capacity to produce the strongest constraints on Galileons.

The most general metric, formed from only $g_{\mu\nu}$ and a scalar field $\phi$ that respects causality and weak equivalence, was shown in Ref.~\cite{Bekenstein:1992pj} to be
\begin{equation}
 \tilde{g}_{\mu\nu}=A(\phi,X)g_{\mu\nu}+B(\phi,X)\partial_\mu\phi\partial_\nu\phi,
\end{equation}
where $X=\frac{1}{2}g_{\mu\nu}\partial_\mu\phi\partial_\nu\phi$.  The first term gives rise to conformal scalars, whose couplings to matter are heavily constrained by various fifth-force experiments.  For our discussion, only the second term, which gives rise to the disformal coupling, is of importance.  This Lagrangian interaction is given by
\begin{equation}
 \mathcal{L}_{dis}=\frac{B(\phi,X)}{2}\partial_\mu\phi\partial_\nu\phi T^{\mu\nu}_{J},
\end{equation}
where $T^{\mu\nu}_{J}$ is the energy-momentum tensor of all matter fields as given in the Jordan frame.  
 
In previous work \cite{Kugo:1999mf,Kaloper:2003yf,Brax:2014vva}, the leading disformal coupling in nonrelativistic systems was shown to be a one-loop quantum effect that gives rise in atoms to an energy-level perturbation
\begin{equation}
\label{eq:de1}
 \delta E=-\frac{3m_im_j}{32\pi^3M^8}\bigg< E\bigg|\frac{1}{r^7}\bigg|E \bigg>,
\end{equation}
where $m_i,m_j$ are the masses of the constituent particles of the bound state (throughout we will use the convention that $m_i\geq m_j$) and $M$ is the Galileon coupling scale.  This potential is sufficiently singular to be divergent for pointlike particles like $e$ and $\mu$.  In the case of muonic hydrogen, $(\mu H)$, it is possible to handwavingly cut off the integral because the proton is composite, and then one may apply a screening mechanism\cite{Brax:2007ak,Dvali:2010jz,deRham:2012az} to prevent the region inside this radius from contributing.  For the standard, noncomposite leptons using any composite scale as a cutoff scale is not possible. Further we will show that in particular the choice of the charge radius as the Galileon coupling scale assumed in Ref.~\cite{Brax:2014zba} is incompatible with what is known about leptonic bound states.  

We begin in Sec.~\ref{sec:rad} with a discussion of how radii are defined in quantum field theory and the complications that arise with associating the Galileon cutoff radius with any other measured quantity.  In Sec.~\ref{sec:2} we explore ways to bound the possible composite scale of leptons, with particular emphasis on the charge radius.  Section~\ref{sec:3} is devoted to deriving constraints on Galileons from Lamb shift measurements  in leptonic systems.  Finally, we conclude in Sec~\ref{sec:4} with remarks on the future of bounds from leptonic systems. 

\section{On radii}
\label{sec:rad}
In Ref.~\cite{Brax:2014zba}, the cutoff radius, $r_i$, was assumed to be approximated by the charge radius of the proton $\sqrt{\langle r^2\rangle}\equiv r_P$ which is derived from hydrogen spectroscopy to be $r_P=0.8758(77)$ fm.  For this value of $r_i$, $M=320$ MeV was found to explain the $r_P$ discrepancy.  This low value for the coupling scale can be reconciled with other constraints by embedding the disformal scalars in a Galileon theory with screening mechanisms caused by higher-order operators\cite{Brax:2014vva}.  With this in mind, we will only consider models of disformal scalars that are embedded in Galileon models.  In this paper, we study how limits on lepton compositeness affect the Galileon explanation for the muon problem.

There are two ways to interpret the assumption $r_i=r_P$.  Reference~\cite{Brax:2014zba} seems to use the charge radius as a proxy for the size of the nuclei that can be used as a cutoff.  This interpretation is troubling because it prevents sensible predictions of the Galileon in other bound states.  Since nuclei can contain neutrons with a negative $\langle r^2\rangle=-0.1149(35)$ fm$^2$, the charge radius of a nuclei can be reduced while the size of the nuclei could increase.  Another troubling part of using the charge radius is that this choice is arbitrary.  A charged particle has a number of radii, each one reflecting a different distribution (e.g. electric charge, weak interaction, neutron density, strange quark density, matter density), and as we will see, assuming any two are (nearly) equal has implications for other particles.  A further complication of this view is that it provides no explanation as to how leptonic bound systems can regulate the divergence of Eq.~(\ref{eq:de1}).

Another way to understand this assumption is that it expresses a relationship between the underlying distributions.  In this paper we investigate whether this relationship can be sustained quantitatively with leptonic bound states.  To begin, formally the charge radius of a particle is defined via the electric form factor,
\begin{align}
 G_E(q^2)=&\int\mathrm{d}^3xe^{i\bm{q}\cdot \bm{x}}\rho(\bm{x})\nonumber\\&=\int\mathrm{d}^3x\left(1+i\bm{q}\cdot\bm{x}+\frac{(\bm{q}\cdot\bm{x})^2}{2}+\cdots\right)\rho(\bm{x})\nonumber\\&=Q_{\rm tot}-\frac{1}{6}|\bm{q}|^2\langle r^2\rangle+\cdots,
\end{align}
where $G_E$ is the electric form factor, $\rho(\bm{x})$ is the charge density, and $Q_{\rm tot}$ is the total charge of the particle.  The standard definition of $\langle r^2\rangle$ is then
\begin{equation}
\label{eq:rc}
 \langle r^2\rangle=-6\frac{dG_E}{dq^2}\bigg|_{q^2=0}.
\end{equation}
In analog to this, we argue that an appropriate definition for $r_i$ should be via a disformal form factor, and therefore would be
\begin{equation}
 r_i^2=-6\frac{dG_{\rm dis}}{dq^2}\bigg|_{q^2=0}.
\end{equation}
By this definition, we see that $r_i$ is related to a disformal density $\rho_{\rm dis}(x)$ that represents the spatial distribution of matter coupling to the Galileons.  Therefore the assumption $r_i=r_P$ is not an arbitrary statement, but is tantamount to saying 
\begin{equation}
\label{eq:rr}
 \int\mathrm{d}^3x|\bm{x}|^2\rho(\bm{x})=\int\mathrm{d}^3x|\bm{x}|^2\rho_{\rm dis}(\bm{x}).
\end{equation}
  Since setting the right-hand side of Eq.~(\ref{eq:rr}) for leptons to zero is unacceptable due to the divergence in Eq.~(\ref{eq:de1}), this implies that leptons must have a charge distribution different from a point particle.  If instead $r_i$ is a property of particles unconnected to their charge radius, then the so-far unobserved lepton charge radius would give no constraint.  This would be in analogy to how the Zemach radius is a property of charged particles arising from the magnetic field distribution, and therefore has no necessary relation to the charge radius.

\section{Compositeness of leptons}
\label{sec:2}
As stated above, the results of Ref.~\cite{Brax:2014zba} relied upon $r_P>0$ in order to cut off the divergences.  Using the measured value from hydrogen or electron scattering experiments, a novel correction to the Lamb shift of muonic hydrogen can mimic a scenario with a smaller $r_P$.  Since Galileons couple to all matter content equally, this interaction should occur in leptonic systems also.  But if leptons are truly pointlike, the potential would lack regularization and would yield unacceptably large corrections to the Lamb shift and $1s-2s$ intervals.  In order to prevent this effect, we are forced to introduce a composite radius. Here, we investigate the strict constraint that $\sqrt{\langle r^2\rangle}=r_i$ (i.e., the charge radius is the composite radius), and a more general constraint on a composite radius.

Since we must demand screening mechanisms to evade other bounds, it is nontrivial to construct a composite scale from Galileons alone; therefore, we compute several different limits on composite lepton radii.
\subsection{Spectroscopy}
Derived constraints from hyperfine splitting (hfs), Lamb shifts, and $1s-2s$ intervals are either directly on the $\langle r^2\rangle$ or on the Zemach radius, $\langle r\rangle_{(2)}$.  The Zemach radius is approximately linearly related to $\sqrt{\langle r^2\rangle}$, with a model-dependent $\mathcal{O}(1)$ coefficient(See Ref.~\cite{Distler:2010zq} for a discussion).  Composite leptons have finite-size contributions similar to those of the proton\cite{Eides:2000xc}.  For the case of an $s$-state energy level, the leading finite-size contribution is known, 
\begin{equation}
 \delta E=\frac{2}{3n^3}\left(Z\alpha\right)^4\mu^3\langle r^2\rangle,
\end{equation}
where $n$ is the principal quantum number, $Z$ is the charge of the particle, and $\mu$ is the reduced mass of the system.  At this order, $p$ states are not affected by $\langle r^2\rangle$, so for the Lamb shift, the contribution is 
\begin{equation}
 \delta E_{\rm Lamb}=\frac{1}{12}\left(Z\alpha\right)^4\mu^3\langle r^2\rangle.
\end{equation}
Furthermore for the $1s-2s$ interval, this contribution yields 
\begin{equation}
 \delta E_{\rm 1s-2s}=\frac{7}{12}\left(Z\alpha\right)^4\mu^3\langle r^2\rangle.
\end{equation}
We can derive stronger limits from the hyperfine splitting (hfs), where the leading-order effect is given in Ref.~\cite{Zemach:1956zz}:
\begin{equation}
 \delta E_{\rm hfs}=-2\left(Z\alpha\right)\mu\langle r\rangle_{(2)}E_F,
\end{equation}
where the Fermi energy is given by
\begin{equation}
 E_F=\frac{8}{3}(Z\alpha)^4(1+a_i)\frac{m_j}{m_i}\left(\frac{\mu}{m_j}\right)^3m_e,
\end{equation}
where $a_i$ is the anomalous magnetic moment of particle $i$.  In addition to the Zemach radius, there is a higher-order contribution directly from $\langle r^2\rangle$, which is found in Ref.~\cite{Karshenboim:1996ew}:
\begin{equation}
 \delta E_{\rm hfs}=\frac{4}{3}\left(Z\alpha\right)^2\ln\left(Z\alpha\right)\mu^2\langle r^2\rangle E_F.
\end{equation}
Applying these expressions to the measured energy spectrum of positronium and muonium states, we can set limits on $\langle r\rangle_{(2)}$ and $\langle r^2\rangle$.  While the Lamb shift and $1s-2s$ constraints in muonium can be applied to either the $e$ or $\mu$ component, the shift to the hfs is not mass symmetric, so these limits apply only to muons.  Table~\ref{tab:radius} is devoted to listing the various constraints, using the experimental and theoretical values found in Table~\ref{tab:data}.  We point out that the Zemach radius constraint in muonium gives a numerical bound very close to that from $a_\mu$, and so in Fig.~\ref{fig:1} we only label the Zemach radius.   

In Table~\ref{tab:radius} we have used $x$ as a stand-in for the percent uncertainty in a measurement of energy shifts in true muonium.  We see that, for even a 10\% measurement of the Lamb shift or hyperfine splitting, true muonium would give competitive limits on the muon's size.
\begin{table}[t]
\begin{center}
 \begin{tabular}{l c c c c}
 \hline\hline
  Atom & $\sqrt{\langle r^2 \rangle}_{1s-2s}$(m) & $\sqrt{\langle r^2 
\rangle}_{\rm Lamb}$(m) &$\sqrt{\langle r^2 \rangle}_{\rm hfs}$(m)&  $\langle r 
\rangle_2$(m)\\
  \hline
  Mu &$4\times10^{-15}$	&$1\times10^{-14}$&$4\times 10^{-18}$ &$1\times 
10^{-18}$\\
  Ps &$7\times10^{-15}$	&$8\times10^{-15}$&$2\times 10^{-16}$ &$2\times 
10^{-15}$\\
  TM &$4\sqrt{x}\times10^{-13}$	&$8\sqrt{x}\times10^{-15}$&$\sqrt{x}\times 
10^{-18}$ &$x\times 10^{-17}$\\
  \hline\hline
 \end{tabular}
\end{center}
 \caption{\label{tab:radius}Constraints on the charge radius and Zemach radius for 
leptonic systems.  For true muonium (TM), $x$ corresponds to the percent precision 
of a future measurement.}
\end{table}

%
\subsection{Anomalous magnetic moments}
In order to solve the muon problem, considering constraints from the anomalous magnetic moments is critical.  From the formalism of Brodsky and Drell \cite{Brodsky:1980zm}, one can use the precisely measured $a_l$ to limit possible substructure in the leptons.  Models of composite leptons can generically give corrections $\Delta a_l\propto m_l/M^*$, where $M^*$ is the scale of new physics. As pointed out in Ref.~\cite{Brodsky:1980zm} though, these models result in a strong fine-tuning to the self-energy of the form $\delta m_l\propto m_l/M^*$.  A more conservative, and perhaps more reasonable, estimate of $a_l$ assumes the existence of a chiral symmetry which causes a cancellation of the linear term in $m_l$, leaving $\Delta a_l\propto \left(\frac{m_l}{M^*}\right)^2$.

If we use the current best limits from experiments compared to theory for the values of $a_l$ \cite{PhysRevD.73.072003,Hanneke:2010au,Aoyama:2014sxa},
\begin{align}
 \Delta a_e=&-9.1(8.2)\times10^{-13},\nonumber\\
 \Delta a_\mu=&\phantom{-}287(80)\times10^{-11},
\end{align}
and again argue that the corresponding mass scales can be interpreted as limits on the radius via $R\approx\hbar/M^*c$, we find flavor-dependent limits on the composite radius of 
\begin{align}
 R_{e}\lesssim &4\times10^{-19}\text{ m},\nonumber\\
 R_{\mu}\approx&1\times10^{-18}\text{ m},
\end{align}
where, because of the discrepancy between theory and experiment, the results for $\mu$ are a preferred scale as opposed to a limit.
The upcoming $(g-2)_\mu$ experiment \cite{Grange:2015fou} anticipates a factor of 4 improvement in the measurement of $a_\mu$, which could improve our limit by a factor of 2.  These limits lack a clear relation to the lepton $\langle r^2\rangle$ but are strong limits on compositeness.  If we seek to explain the complete muon problem with Galileons, a precision constraint from $a_\mu$ would be essential to derive.  

\section{Limits on Galileons}
\label{sec:3}
\begin{figure}[ht]
 \begin{center}
  \includegraphics[width=\linewidth]{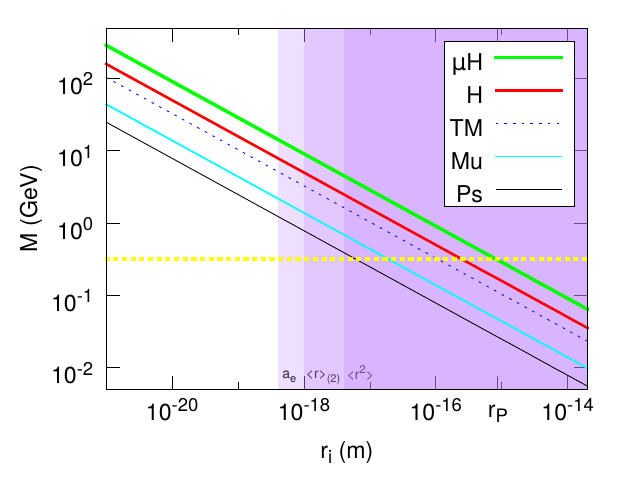}
 \end{center}
 \caption{\label{fig:1} Allowed ($r_i$,$M$) parameter space from various atomic systems.  The solid lines indicate current 
constraints from Eq.~(\ref{eq:gals}), while the dashed line for true muonium indicates a future measurement with 50\% precision.  The parameter space below the lines is excluded.  The horizontal line is the value of $M$ to explain the $r_P$ discrepancy found in 
Ref.~\cite{Brax:2014zba}.  Shaded regions indicate excluded values of 
leptonic radii for $\mu,e$ and hfs measurements, as 
explained in text and summarized in Table~\ref{tab:radius}.  To aid comparisons, we have indicated $r_P$ on the $x$-axis.}
\end{figure}

It was shown in Ref.~\cite{Brax:2014zba} that the leading correction to the 
Lamb shift of an atomic system due to a Galileon is
\begin{equation}
\label{eq:gals}
 \delta 
E_{\rm Lamb}=\frac{3}{248\pi^3}\left(\frac{Z}{a_0}\right)^3\frac{m_i 
m_j}{M^8r_i^4}\left[1-\frac{1}{6}\left(\frac{Z}{a_0}\right)^2r_i^2\right],
\end{equation}
where $a_0$ is the Bohr radius of the atom and $r_i$ is the radius of particle $i$ at which the divergence is cut off.  To derive constraints from this equation, we compute the $1\sigma$ value of $\delta E_{\rm Lamb}$ for each leptonic system found in Table~\ref{tab:data}.  Our procedure for this is to first combine the errors in quadrature, and then sum this with the observed value of $\delta E_{\rm Lamb}$. Equating these results to Eq.~(\ref{eq:gals}), parts of the parameter space of the $r_i$ and $M$ below the lines in Fig.~\ref{fig:1} are excluded at $1\sigma$.  We find for any fixed value of $r_i$, hydrogenic systems place stronger limits than leptonic ones on the value of $M$.  While this result at first seems disappointing, it is important to remember that in hydrogenic systems one previously assumed that $r_i=r_p$.  In contrast, lepton systems can take on any values of $r_i$, because no composite scale has been measured.  

Using the constraints on the various radii derived in Sec.~\ref{sec:2}, we see that leptonic 
systems require the scale of $M$ to be higher than that preferred by $\mu H$. Currently, positronium energy shifts alone are not well enough known to competitively limit the charge or Zemach radius, so we have not plotted them.  But combining positronium constraints from Eq.~(\ref{eq:gals}) with the $a_e$, we obtain the bound $M_{a}>1.2$ GeV.  For muonium, all limits on $r_i$ are superior bounds on $M$ to those from the muonic hydrogen which require $r_i=r_P$.  These bounds vary from $M_{\sqrt{\langle r^2\rangle}}>0.67$ GeV to $M_{\rm a}>1.33$ GeV.  In particular, we emphasize that the constraint from the charge radius of the muon excludes $M=320$ MeV.  We therefore conclude that leptonic systems rule out Galileons as the explanation of the muon problem if the radius is related to the charge or Zemach radius.

Further, we notice that a fiducial measurement of the Lamb shift in true muonium with only 50\% precision would improve our limit on the scale $M$ by a factor of 2.  In fact, due to the weak $M\propto E^{-1/8}$ scaling of our constraints, a mere measurement of the \textit{existence} of a Lamb shift in true muonium would likely provide the strongest limit on Galileons from leptonic physics.
  
Having seen that requiring $\sqrt{\langle r^2\rangle_{em}}=r_i$ for leptons would rule out the preferred value of $M$ from $(\mu H)$, we can ask whether this assumption is necessary. It is perhaps not surprising that these two scales must be different.  $\sqrt{\langle r^2\rangle_{em}}$ is a property of particles determined by their charge distribution, while $r_i$ should in principle be defined in a similar way by the field distribution that couples to Galileons.  Since the latter couples not only to quarks and leptons, but photons, gluons, W's and Z's, the two distributions should differ. This is analogous to how the magnetic distribution of the proton results in the Zemach radius being different from the charge radius.  Allowing $r_i\neq r_P$ for the proton allows the muon problem to be solved for any value along the $(\mu H)$ line in $(r_i,M)$ space not excluded by limits on $M$ from other sources, and since these are currently the strong compositeness constraints, an unobserved lepton $r_i$ is acceptable.
\begin{table*}[t]
\begin{center}
 \begin{tabular}{l c c c c c c}
 \hline\hline
  Atom & Obs. & $\Delta E_{exp} $(MHz)& Exp. Ref.& $\Delta E_{theory}$(MHz) & Theory Ref.&$\delta 
E$(MHz)\\
  \hline
  Mu & Lamb 
&1042(23)&\cite{Woodle:1990ky}&1047.490(300)&\cite{Bhatt:1987zz,Erickson:1988zz,
Lautrup:1971jf}	&-5.5(230)(0.3)	\\
   & $1s-2s$ &2455528941.0(9.8)&\cite{Meyer:1999cx} 
&2455528935.4(14)&\cite{Karshenboim:1996bg,Pachucki:1996jw,Karshenboim:1997zu}	
&5.6(98)(14)	\\
    &hfs&4463.302765(53)&\cite{Liu:1999iz} 
&4463.30288(55)&\cite{Karshenboim:2006ht}&-0.000115(53)(55)	\\
  Ps & Lamb & 13012.42(67)(154)&\cite{PhysRevLett.71.2887} 
&13012.41(9)&\cite{PhysRevA.60.2792}	&-0.01(67)(154)(9)	\\
   & $1s-2s$ &1233607216.4(32)&\cite{PhysRevLett.70.1397}	
&1233607222.2(6)&\cite{PhysRevA.60.2792}	&-5.8(34)(6)	\\
   &hfs&203389.10(74)&\cite{PhysRevA.30.1331} 
&203392.411(60)&\cite{marcu2011ultrasoft,Adkins:2014dva,Baker:2014sua,
Eides:2014nga,Adkins:2014xoa}	&-3.31(74)(48)(6)	\\
  &hfs&203394.2(16)(13)&\cite{Ishida:2013waa}&203392.411(60)&\cite{marcu2011ultrasoft,Adkins:2014dva,Baker:2014sua,
Eides:2014nga,Adkins:2014xoa}&1.78(160)(130)(6)	
\\
  TM & Lamb &$\cdots$& $\cdots$ & 1.35(5)$\times 10^{7}$&\cite{Jentschura:1997tv}	  
&6.8(5)$\times 10^{6}$	\\
  & $1s-2s$ &$\cdots$ &$\cdots$ & 2.55(5)$\times 10^{11}$&\cite{Jentschura:1997tv}	& 
1.27(5)$\times 10^{11}$\\
    & hfs &$\cdots$ &$\cdots$ & 
42330577(800)(1200)&\cite{PhysRevD.91.073008}&21165288(800)(1200)	\\
  \hline\hline
 \end{tabular}
\end{center}
 \caption{\label{tab:data}Experimental and theoretical values for the necessary 
energy shifts in leptonic systems.  For the case of true muonium, we have used 
the representative value of 50\% of the theoretical values for $\delta E$}
\end{table*}

\subsection{Limits from perturbativity}
The condition that the disformal scalars avoid other constraints required us to embed it in a model with chameleon\cite{Khoury:2003aq,Khoury:2003rn} and Galileon properties.  In Ref.~\cite{Brax:2014zba}, the full action was given by
\begin{align}
S=&\int d^4x \sqrt{-g}\bigg(\frac{R}{16\pi G_N} -\frac{1}{2} (\partial \phi)^2 -\frac{1}{\Lambda^3}\Box\phi (\partial \phi)^2\nonumber\\&-V(\phi)+ \frac{1}{M^4} \partial_\mu\phi\partial_\nu\phi T_J^{\mu\nu}\bigg)
 \; + S_m(\psi_i, A(\phi)g_{\mu\nu})\;,
\end{align}
where we must introduce the suppression scale $\Lambda$ to control the self-interactions of the Galileons.  Within this model, an upper limit on $\Lambda$ can be set by requiring perturbative unitarity to have not been violated up to currently observed energies. Reference~\cite{Brax:2014zba} finds that using LEP data, which constraints unitarity violations up to 200 GeV,
\begin{equation}
 \Lambda^3\leq \frac{\beta}{2\pi m_{Pl}}\left(\frac{8\pi M^4}{\sqrt{2}m_e}\right)^{4/3},
\end{equation}
where $\beta$ is an $\mathcal{O}(1)$ Galileon coupling.  Using this bound, we have a relation between $\Lambda$ and $M$ of
\begin{equation}
\label{eq:1}
 \Lambda\leq 6\text{ keV}\left(\frac{M}{320\text{ MeV}}\right)^{16/9}.
\end{equation}

Additionally, we have assumed that the Galileon is well approximated by a free scalar field at least down to the scale of the composite radius.  In order to justify this assumption, it was found in Ref.~\cite{Brax:2014zba} that
\begin{equation}
\label{eq:2}
\Lambda \gtrsim \frac{1}{r_i}\left(\frac{\beta m_i}{m_{Pl}}\right)^{1/3}.
\end{equation}
Putting Eqs.~(\ref{eq:1}) and~(\ref{eq:2}) together, we can obtain additional constraints on the space of $(r_i,M)$
\begin{equation}
 \frac{1}{r_i}\left(\frac{\beta m_i}{m_{Pl}}\right)^{1/3}\leq 6\text{ keV}\left(\frac{M}{320\text{ MeV}}\right)^{16/9}
\end{equation}
Unlike the strict limits found in the previous section, this constraint only applies if we demand trustworthy perturbative solutions from Galileons.  With this caveat, for both muons and electrons this bound is violated if $M=320$ MeV, again ruling out the assumption that $r_i=\sqrt{\langle r^2\rangle}$. 

\section{Summary and Conclusions}
\label{sec:4}
In this paper, we have shown how leptonic systems offer competitive constraints 
on the scale $M$ of Galileons.  From the nonobservation of a lepton charge or Zemach radius, we can be confident that either $\sqrt{\langle r^2\rangle}\neq r_i$ or the Galileon scale must be $M>1.33$ GeV.  This result would be competitive with collider and astrophysical constraints.  Going beyond the assumption that the two radii should be related, the Galileon model is still viable for solving the muon problem.

Looking forward, the improvement of these atomic constraints is possible.  Understanding the true finite-size effects of Galileons requires going beyond the crude estimate that $\sqrt{\langle r^2\rangle_{em}}=r_i$, and accurately determining their relation.  There are a number of ways this could potentially be done.  The simplest theoretical, but relatively difficult experimentally, approach is to note that since $s$ and $p$ states have the same dependence on $M$ but a different dependence on $r_i$, measurements of the ratio of the absolute energy level of each state would give a limit on $r_i$ alone.  Another possibility is that if one could construct additional Galileon observable effects in $a_l$ or the hfs a similar difference in dependence might arise, allowing us to leverage those precision experiments further.  One could consider improving the experimental precision, but since $M\propto r_i^{-1/2}\delta E^{-1/8}$, many orders of magnitude improvement in experiments and theory will be required to 
improve these limits.  The exception  is the potential measurement of the spectrum in true muonium.  Any Lamb shift measurement in true muonium will improve these limits, and an ambitious part-per-million-level measurement could completely exclude Galileons. Finally, we note that in positronium and true muonium, the existence of an annihilation channel $ll\rightarrow \phi \rightarrow ll$ allows for potentially stronger limits from energy shifts or decay-rate limits from processes like $(\mu^+\mu^-)\rightarrow e^+e^-$. 

\begin{acknowledgments}
The author would like to thank Richard Lebed and Francis Duplessis for helpful comments while developing this work.  This work was supported by the National Science Foundation under Grants No. PHY-1068286 and PHY-1403891.
\end{acknowledgments}
\bibliographystyle{apsrev4-1}
\bibliography{wise}
\end{document}